\documentclass[runningheads,a4paper]{llncs}

\usepackage{amsmath,amssymb}
\usepackage{graphicx} 
\usepackage{subfigure} 
\usepackage{multirow}
\usepackage[]{natbib}

\usepackage{algorithm}
\usepackage{algorithmic}
\usepackage{hyperref}

\usepackage[T1]{fontenc}

\def\tprime{t^{'}}
\def\iter{\textrm{iter}}
\usepackage{amssymb}
\usepackage{graphicx}

\usepackage{url}

\begin{document}

\mainmatter  

\title{Learning About Meetings}

\titlerunning{Learning About Meetings}
\author{Been Kim 
\and Cynthia Rudin}
\authorrunning{Kim and Rudin}
\institute{Massachusetts Institute of Technology\\
\url{beenkim@csail.mit.edu, rudin@mit.edu}}

\maketitle

\begin{abstract}
Most people participate in meetings almost every day, multiple times a day.
 The study of meetings is important, but also challenging, as it requires an
 understanding of social signals and complex interpersonal dynamics. Our aim 
this work is to use a data-driven approach to the science of meetings. We provide
 tentative evidence that: i) it is possible to automatically detect when during 
the meeting a key decision is taking place, from analyzing only the local dialogue
 acts, ii) there are common patterns in the way social dialogue acts are interspersed 
throughout a meeting, iii) 
at the time key decisions are made, the amount of time left in the meeting can be predicted from the amount of time that has passed, iv) 
it is  often possible to predict whether a proposal during a meeting will be accepted or
 rejected based entirely on the language (the set of persuasive words) used by the speaker.
\end{abstract}

\section{Introduction}
\textit{
``A meeting is indispensable when you don't want to get
anything done.'' \citep{kayser1990mining}
}

In the United States alone, an estimated 11 million meetings take place during a typical work day \citep{meetingNumber}. 
Managers typically spend between a quarter and three-quarters of their time in meetings \citep{mackenzie2009time}, and approximately 97\% of 
workers have reported in a large-scale study \citep{hall1994americans} that to do their best work, collaboration is essential. Most of us work 
directly with others every day, and want to be useful participants in meetings. Meeting analysis (the science of meetings) can
 potentially help us understand various aspects of meetings, find ways to allow us to be more effective participants in our meetings,
 and help us to create automated tools for meeting assistance. Meeting dynamics can be complex; often proposals are implicitly 
communicated and accepted \citep{Eugenio99theagreement}. Despite the plethora of meetings that happen each day,
 and despite a body of work on meeting analysis \citep[e.g., see][for a review]{romano2001meeting}, 
we still cannot claim that we understand the topic of meetings well enough that these studies have led to quantifiable 
improvements in the overall quality of meetings, nor useful guidelines for more productive meetings. Perhaps if we step back and consider a
 basic science approach to meeting analysis, we would quantitatively uncover aspects of meetings that could eventually lead to true improvements.

In this work, we develop and use predictive modeling tools and descriptive statistics in order to provide a data-driven approach to the scientific study of meetings.
We provide preliminary answers to several key questions: i) Can we automatically detect when 
the main decisions are made during the meeting, based only on the frequency and type of dialogue acts in a given period of time? In other words, based on the types of utterances people are making, can we determine whether the most important part of the meeting is occurring? 
 ii) Is there any type of pattern of dialogue common to most or all meetings? We would like to know whether
 there is a ``quintessential'' pattern of dialogue that we can identify. 
iii)
 How long is a meeting going to last, 
in terms of ``wrap-up'' time, beyond the time that the main decisions are made? Sometimes this wrap-up time is substantial, 
and the meeting extends well beyond when the main decisions are made. iv) 
Can we predict whether a proposal made during 
a meeting will be accepted or rejected based entirely on the language (the set of persuasive words) used by the speaker? 
There have been many studies and commentaries focused on ``persuasive words,'' but do those persuasive words truly correlate with successful 
outcomes during a meeting? Some of these 
questions can be answered using current machine learning approaches, but others cannot. For finding patterns in dialogue acts 
that are common to most/all meetings, we design an algorithm to learn a sparse representation of a meeting 
as a graph of dialogue acts, which yields insight about the way social dialogue acts intersperse with work-related dialogue acts.

In our study, we use the most extensive annotated corpus of meetings data in existence, originating from the AMI (Augmented multi-party interaction) 
project \citep{Mccowan05theami}. This dataset contains a large number of annotated meetings (over 12,000 human-labeled annotations). 
In total, there are 108,947 dialogue acts in total number of 95 meetings, 
and 26,825 adjacency pairs (explained below). 
This corpus is derived from a series of real meetings, controlled in the sense that each meeting has four participants who work as a
 team to compose a new design for a new remote control. 
 Each participant takes a role, 
either project manager, marketing expert, industrial designer or interface designer. The participants are given training for their roles 
at the beginning of the task. Documents used to train participants and annotators are publicly available. The interaction between participants
 is unstructured, and each person freely chooses their own style of interaction. The length of meetings ranges from approximately 10 minutes to 45 minutes,
which overlaps with the most common lengths of meetings \citep[discussed by][]{romano2001meeting}. Here we provide detail on the annotations provided with the corpus.

\textit{Dialogue Acts}:  
A dialogue act marks a characteristic of an utterance, representing the intention or 
the role of the utterance. It is possible to use dialogue act data to 
predict characteristics of future dialogue \citep{nagata1994first}, or to do
automatic dialogue act tagging \citep{stolcke2000dialogue,ji2005dialog}.
Dialogue acts include questions, statements, suggestions, assessment of suggestions (positive, negative or neutral)
and social acts. Each sentence is often divided into pieces and tagged with 
different dialogues acts.\footnote{More details of definition of dialogue acts can be found 
in \citep{Mccowan05theami}.} 
A sequence of dialogue acts looks like this:\\
\texttt{ 
A: Suggestion\\
B: Commenting on A's suggestion \\
C: Asking questions\\
A: Answering C's question\\
B: Accepting A's suggestion
}

\textit{Decision summary}: A summary of decisions made in a meeting and related
dialogue acts.

 \textit{Discussion}: A set of dialogue acts and their time stamps
that support or reflect decisions described in the decision summary annotation. 
 
\textit{Adjacency pairs}: An adjacency pair encodes the relationship between two 
dialogue acts. It represents the reaction to the first dialogue act that is expressed
 within the second dialogue act; e.g., the first dialogue act can be a suggestion
 and the second dialogue act can be a positive or negative response to it, 
as demonstrated in the sequence of dialogue acts above. 

We note that meetings can have many purposes, \citep[e.g., see][]{romano2001meeting}, however in this work we study only 
meetings where the purpose is to make a group judgment or decision  (rather than for instance, to ensure that everyone understands, or to explore new ideas and concepts).

Our specific goal in this work is to contribute insights to the new scientific field of meeting analysis. We provide tentative answers to the above questions 
that can be used for constructing other studies. We do not claim that our results definitively answer these questions, only that they yield hypotheses that can be tested more thoroughly through other surveys. Most medical (and more generally, scientific) studies make or verify hypotheses based on a database or patient study. Hypotheses based on these single studies are sometimes first steps in answering important questions - this is a study of that kind, and we believe our results can be tested and built upon.

Our more general goal is to show that ML can be useful for exploratory study of meetings, including and beyond the questions studied here.
 This paper is not focused on a specific machine learning technique per se, its goal is to show that ML methods can be applied for exploratory 
scientific analysis of data in this domain, leading towards the eventual goal of increasing meeting productivity.
We start with a question for which a simple application of machine learning tools provides direct insight.

\section{Question i: Can we automatically detect when key decisions are made?
\label{q1}}
If we can learn the statistics of dialogue acts around the time when important decisions are about to be made or being made,
 we can potentially detect the critical time window of a meeting. One can imagine using this information in several ways, 
for instance, to know at what point in the meeting to pay more attention (or to join the meeting in the first place), 
or to use this as part of a historical meetings database in order to fast forward the meeting's recording to the most important parts.

In our setup, each feature vector for the detection problem is a time-shifted bag-of-dialogue-acts. The feature vector for 
a specific timeframe is defined by the counts of different dialogue acts (e.g., one example has 3 suggestions, 2 positive responses and 10
information exchanges). Using this representation means that the results do not depend on specific keywords within the meeting dialogue; this 
potentially allows our results to hold independently of the specific type and purpose of the meeting being held. 
Among dialogue acts, we use only a subset that are known to be relevant to the group decision making process \citep{Eugenio99theagreement},
namely: action directive, offer, accept, reject, info-request, and information. Action directives represent all elicit
forms of dialogue acts --- dialogue acts that require actions from hearers. 
By limiting to these dialogue acts, we are now working with a total of 53,079 dialogue acts 
in the corpus. 

Using this definition, a timeframe of dialogue becomes one example, represented by a 6-dimensional vector, where each element is
the count of a particular dialogue act within the specified timeframe. The labels are 1 if the decisions
are annotated as being made in that timeframe of interest and 0 otherwise. We considered timeframes with average size 
70 timestamps (roughly 5 minutes). This choice of timeframe comes from the minimum meeting time considered in the research done
on meeting profiles by \cite{panko1995meeting}.

We applied supervised and unsupervised classification algorithms to predict whether important decisions are being made within a given block of dialogue. 
The data were divided into 15 folds, and each fold was used in turn as
the test set. The supervised algorithms were SVM with radial basis function (RBF) kernels, SVM with linear kernels, logistic regression, and Na\"ive Bayes with a Gaussian 
density assumption on each class and feature. 
For the unsupervised algorithms, which (purposely) discard the annotations, we used EM with Gaussian Mixture Models and Kmeans.
From the resulting two clusters, we chose the better of the two possible labelings. 
The AUC, precision, recall, and F-measure for each algorithm's performance on the testing folds are reported in Table \ref{tab:results_for_A}. The 
sample mean and sample standard deviation of AUC values over the test folds are reported within the second column of the table.
 
\begin{table}
\caption{Results for predicting key decision times\label{tab:results_for_A}}
\vskip 0.15in
\begin{center}
 \begin{sc}
\begin{tabular}{ | l | l| l| l |l |l |}
\hline
   \bf{Method}	        &\bf{AUC $\pm$ Std.  }\;\;\;\;& \bf{Precision} & \bf{Recall} & \bf{F-measure} \\ \hline 
SVM-Linear&    		0.87 $\pm$ 0.05 & 0.89 &  0.87 &  0.88 \\ 
Logistic regression\;\;&    0.87 $\pm$  0.05 & 0.70 &  0.56 &  0.62 \\ 
SVM-RBF&	 	0.86 $\pm$  0.06 & 0.86 &  0.95 &  0.90 \\ 
EM-GMM&      		0.57 $\pm$  0.06 & 0.68 &  0.48 &  0.56 \\ 
NB-Gaussian &    	0.56 $\pm$  0.10 & 0.88 &  0.81 &  0.84 \\ 
Kmeans&	  		0.48 $\pm$  0.24 & 0.69 &  0.34 &  0.45 \\ 
\hline
   \end{tabular}
\end{sc}
 \end{center}
\vskip -0.1in
\end{table}


The prediction quality, with respect to the AUC, is very similar for SVM-Linear, Logistic Regression, and 
SVM-RBF. All three methods show high AUC values around 0.86 or 0.87. The three other methods do not perform 
nearly as well. Na\"ive Bayes has very strong independence assumptions that are clearly violated here. EM-GMM 
is a non-convex approach that has a tendency to get stuck in local minima. Both EM-GMM and K-Means are unsupervised, so 
they do not get the benefit of being trained on human-labeled data. We remark that the predictive performance given by some of these 
algorithms is quite high given that the imbalance ratio is only 3 to 1 (three ``no decision made'' examples for each ``decision made'' example). Logistic 
regression performed well with respect to the AUC, but not with respect to the other measures. AUC is a rank 
statistic, whereas the other measures are relative to a decision boundary. Thus, if the decision boundary is in the 
wrong place, precision, recall, and F-measure suffer, regardless of whether the positives are generally given higher scores than the negatives. 
(There is an in-depth discussion by \cite{ErtekinRu11} on logistic regression's performance with respect to the AUC.) The opposite is true 
for Na\"ive Bayes, where the decision boundary seems to be in the right place, leading to good classification 
performance, but the relative placement of positives and negatives within the classes lead to rank statistics that are not at the level of the other algorithms. This could potentially be due to the choice of scoring measure used for the AUC, as the choice of scoring measure is not unique for Na\"ive Bayes. We chose $P(y=1)\prod_j \hat{P}(x^j|y=1)$, where the empirical probability is computed over the training set. (Here $y=1$ indicates decisions being made, and $x^j$ is the $j^\textrm{th}$ feature value.)
It is possible that even if an example scored highly for being a member of the positive class, it could score even more highly for being a member of the negative class, and thus be classified correctly. Thus, care should be taken in general when judging Na\"ive Bayes by the AUC; in this case, quantities computed with respect to the decision boundary (true positives, false positives, etc.) are more natural than rank statistics.


%

\begin{table}
\caption{Feature ranking using SVM coefficient \label{tab:q1ranking_svm}}
\begin{center}
    \begin{tabular}{| p{1.5cm} | p{3.5cm}  | p{2.5cm} |}
    \hline
\bf{Ranking} & \bf{Dialogue Acts} & \bf{$\lambda$ $\pm$ Std.}  \\ \hline
1& Information & 0.30 $\pm$0.031\\
2& Information Request & 0.11 $\pm$ 0.03\\
3& Offer& -0.0076 $\pm$ 0.04\\
4& Action-directive& -0.0662 $\pm$0.04\\
5& Reject& -0.20 $\pm$0.03\\
6& Accept& -0.27 $\pm$0.02\\
\hline
    \end{tabular}
\end{center}
\end{table}
We can use the SVM coefficients to understand the distribution of dialogue acts during the important parts 
of the meeting. As it turns out, the important parts of the meeting are characterized mostly by information 
and information request dialogue acts, and very few offers, rejections, or acceptances. This is shown 
in Table \ref{tab:q1ranking_svm}. We hypothesize that at the important parts of the meeting, when the decisions
 have been narrowed down and few choices remain, the meeting participants would like to ensure that they have all
 the relevant information necessary to make the decision, and that the outcome will fit within all of their constraints.

%

%

%
%
%
%
%

This small experiment has implications for the practical use of machine learning
 for automated meeting recording and assistance software. First, it is possible
 to obtain at least 0.87 AUC in detecting the time frame when decisions are 
made. An algorithm with this level of fidelity could be useful, for instance, 
in searching through large quantities of meeting data automatically (rather than 
manually scrolling through each meeting). 
One could envision having software create an alert that key decisions are being made, so that 
upper-level management can then choose to join the meeting. Or the software could inform secretarial 
staff of an estimate for when the meeting is done in order to facilitate schedule planning. (To do this, however, it might be useful to incorporate knowledge from Section \ref{q3} about the expected length of the wrap-up time.)


\section{Question ii: Is there a pattern of interactions within a meeting?\label{q2}}
There is often a mixture of social and work-related utterances during a meeting, and it is not obvious how the two interact. In particular, we would like to study the way in which social acts (positive or negative) interact with work-related acts (i.e., acceptance or rejection of proposals). More abstractly, we would like to know if there is a ``quintessential representation'' of interactions within a meeting, where a representation is a directed graph of dialogue acts. If there is such a representation, we would like to learn it directly from data.
To do this, we will present a discrete optimization approach that uses the notion of ``template instantiation.'' The optimization will be performed via simulated annealing. First, let us formally define the problem of template discovery.

\subsection{Formalization of the Problem of Template Discovery}
\begin{figure}
\begin{center}
 \includegraphics[width=3.2in]{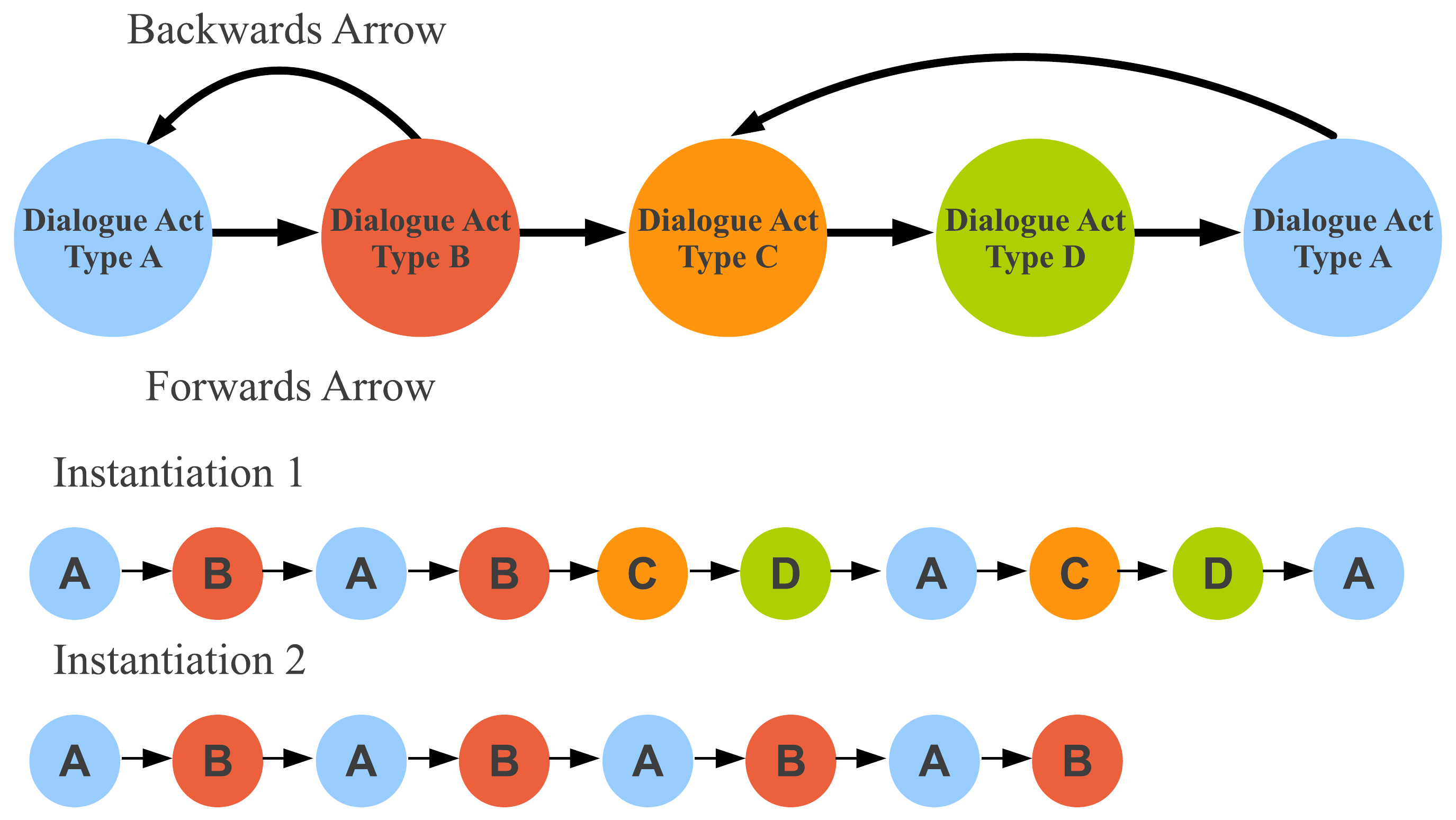}
 \caption{\label{template_enum} Example template instantiations. The template is at the top, and two instantiations are below.}
\end{center}
\end{figure}

We define a \textit{template} as a graph, where each node takes a value in the set of dialogue acts, 
and the graph has directed edges from left to right, and additional backwards edges. Formally, let $\Lambda$ be the 
set of possible dialogue acts ($\Lambda^n$ is a string of length $n$). 
Specifically, define the set $\mathcal{D}$ as 
\[\mathcal{D}: ( n \in \{ 1, \cdots, L\} ) \times \Lambda^n \times \Pi_{B,n} \]
so that $\mathcal{D}$ is a set of templates with length of size $n \leq L$, where if the template is of size $n$, there are $n$
dialogue acts. There are forward edges between neighboring nodes (i.e. dialogue acts). $\Pi_{B,n}$ is the set of all possible
backward arrows for the template, containing at most $B$ backward arrows. One can represent $\Pi_{B,n}$ as the set of $n \times n$ lower diagonal
binary matrices, 1's indicating backward arrows, with at most $B$ values that are 1.
Let $X$ be the set of meetings, which are strings consisting of an arbitrary number of elements of $\Lambda$.
Define the loss function $l: \mathcal{D} \times X \to \mathbb{Z_{+}}$ as 
\[l(t,x) := \displaystyle\min_{t_j \in \textrm{instan}(t)} [\textrm{dist}(t_j, x)] \]
where dist is edit distance, $\textrm{dist}:\Lambda^{n_1} \times \Lambda^{n_2} \to \mathbb{Z_+}$ for strings of lengths $n_1$ and $n_2$. We define $\textrm{instan}(t)$ as the set of \textit{template instantiations} $t_j$ of the template $t$, where a template instantiation is a path through the graph, beginning or ending anywhere within the graph. The path is a sequence of elements from $\Lambda$. Consider for instance the example template at the top of Figure \ref{template_enum}. Two example template instantiations are provided just below that. The edit distance between two strings is the minimum number of insertions, deletions, and/or substitutions to turn one string into the other. Each meeting $x$ is a string of  dialogue acts, and each template instantiation $t_j$ is also a string of dialogue acts, so the edit distance is well-defined.


We assume that we are given meetings $x_1, \cdots, x_m$ drawn independently from an unknown probability distribution $\mu$ over possible meetings $X$.
Define empirical risk $R^{\textrm{emp}}(t):\mathcal{D} \to \mathbb{R_+}$ and true risk $R^{\textrm{true}}(t): \mathcal{D} \to \mathbb{R_+}$ as: 
\[
R^{\textrm{emp}}(t):= \displaystyle\frac{1}{m} \displaystyle\sum^m_{i=1} l(t, x_i) \textrm{  and  } R^{\textrm{true}}(t) := \mathbb{E}_{x\sim\mu} [ l(t, x) ].
\]
We can bound the difference between the empirical risk and true risk for the problem of template discovery as follows:
\begin{theorem}
For all $\delta > 0$, with probability at least $1-\delta$ with respect to the independent random draw of meetings $x \sim \mu$, $x_i \sim \mu$, $i=1,...,m$,
and for all templates $t$ in $\mathcal{D}$ with length at most $L$ and with number of backwards arrows at most $B$:
\[R^{\text{\rm true}}(t)\leq R^{\text{\rm emp}}(t) + \sqrt{\frac{\log \displaystyle\sum_{n=0}^{L} \left[|\Lambda|^n \displaystyle\sum_{b=0}^{\min(B,n)} {n(n-1)/2 \choose b}\right] + \log \displaystyle\frac{1}{\delta}}{2m}},\]
where $|\Lambda|$ is the number of elements in the set $\Lambda$. 
\end{theorem}
The proof is in the Appendix. 

This framework for macro-pattern discovery leads naturally to an algorithm for finding templates in real data, which is to minimize the empirical risk, regularized by the number of backwards arrows and the length of the template.

\subsection{Macro-Patterns in Meetings}
Let us explain the data processing. We selected the annotations that were meaningful in this context, namely: socially positive act, 
socially negative act, negative assessment act and positive assessment act. This allows us only to focus on assessments (either social or work related), as they may have a generally more powerful effect on the trends in the conversation than other dialogue acts. That is, the assessments create a ``macro-pattern'' within the meeting that we want to learn.
Repeated dialogue acts by the same person were counted as a single dialogue act.
The selected data contain 12,309 dialogue acts, on average 130 acts per meeting.

We would like to know if all or most meetings have something in common, in terms of the pattern of dialogue acts throughout the meeting (e.g., a smaller sub-conversation in which certain dialogue acts alternate, followed by a shift to a different sub-conversation, and so on).
If this were the case, a template for meetings could be a directed graph, where each meeting approximately follows a (possibly repetitive) path through the graph. Time might loosely follow from left to right. 
We thus learn the directed graph by optimizing the following regularized risk functional using discrete optimization:
\begin{eqnarray}\label{eq:optfunction}
\lefteqn{F(\textrm{template } t)} \nonumber \\
&=&\frac{1}{m}
\sum_{\textrm{meetings }i} \min_{t_j\in\textrm{instan}(t)}
[\textrm{dist}(t_j, \textrm{meeting } i)] 
+  C_1\textrm{length}(t)+C_2\textrm{backw}(t),
\end{eqnarray}
which is a regularized version of the empirical risk defined above. Intuitively, (\ref{eq:optfunction}) characterizes how well the set of meetings match the template using the first term, and the other two terms are regularization terms that force the template to be sparser and simpler, with most of the edges pointing forwards in time. The smaller templates have the advantage of being more interpretable, also according to the bound above, smaller templates encourage better generalization. The length$(t)$  is the number of nodes in template $t$. The value of backw$(t)$ is the number of backwards directed edges in the template graph. We chose $C_1=1$ and $C_2=0.1$ to force fewer backwards edges than forwards edges, meaning a graph that follows more linearly in time. 

Template instantiations always follow paths that exist within the template, whereas a real meeting will likely never exactly follow a template (unless the template is overly complex, in which case the algorithm has wildly overfit the data, which is prevented by validation). 
Our calculation for $\displaystyle\min_{t_j\in\textrm{instan}(t)}[\textrm{dist}(t_j, \textrm{meeting } i)]$ in the first term of  (\ref{eq:optfunction}) is approximate, in the sense that the min over all templates is calculated over all template instantiations that are approximately the same length as the meeting $i$, rather than over all instantiations. As it is likely the minimum would be achieved at an instantiation with approximately the same length as the meeting, this is a reasonable assumption to make in order to make the calculation more tractable.

We optimize (\ref{eq:optfunction}) over templates using simulated annealing, as shown in Algorithm \ref{alg:sa}. Simulated annealing probabilistically decides whether it will move to a neighboring state or stay at the current state at each iteration. 
The neighborhood is defined as a set of templates that are edit distance 1 away from the current template under allowable operations. The allowable operations include insertion of a new node between any two nodes or at the beginning or end of the template, deletion of a node, insertion or deletion of backwards directed edges. For the proposal distribution for simulated annealing, each operation is randomly chosen with uniform probability. 
The acceptance probability of the new template is 1  
if the proposed objective function value is less than the current value.
If the proposed value is larger than the current value, the acceptance function accepts 
with probability
$\exp \left( -\Delta F/T \right) $
where $\Delta F$ represents the difference in objective function value, namely the proposed function value minus the current function value,
and $T$ is the 
current temperature 
(lines \ref{alg:accept_jump1}-\ref{alg:accept_jump2} in Algorithm \ref{alg:sa}).
The annealing schedule is $T=T_0\cdot 0.95^k$, where
$k=800$ is the annealing parameter and $T_0=1000$ is the initial temperature.
After every $k$ accepted steps, 
we restarted the optimization starting at the best point so far, resetting the temperature to its initial value. The maximum iteration number was set at 4000.

\begin{algorithm}[th!]
   \caption{Discrete optimization algorithm for learning meeting templates\label{alg:sa}}
\begin{algorithmic}[1]
   \STATE {{\bfseries Input:} Starting template $t_0$, $m$ meetings, constants $C_1$, $C_2$, initial temperature $T_0$.}
   \STATE {Initialize $t = t_{0}$, $t_{\textrm{best}}=t_0$, $\iter=0$, $k = 800$, set $F(t_o)$.}
   \WHILE{$ (!\mbox{converged}) \cap (\iter < \mbox{Maximum Iteration})$}
   \IF{mod$(\iter)=k$}
       \STATE{$ T \leftarrow T_0$ where $T_0$ is initial temperature, $t\leftarrow t_{\textrm{best}}$}
   \ENDIF
       \STATE{$ N_t \leftarrow $ get\_neighborhood\_templates($t$)}
   \STATE{ choose  $\tprime\in N_t$ uniformly at random}
     \STATE{length$(\tprime)$ $\leftarrow$ number of nodes in template $\tprime$}
     \STATE{backw$(\tprime)$ $\leftarrow$ number of backwards edges in template $\tprime$}
   \FOR{all meetings $i\leq m$}
   \STATE{instan$(\tprime)$ $\leftarrow$ set of template instantiations (sufficient to use only ones with length $\approx$ length of meeting $i$)} \label{alg:template_enum}
  \FOR{ all instantiations $t_j\in$ instan$(\tprime)$}  \label{alg:cal_edit_dist1}
  \STATE{ $d_{ij}^{\tprime}=\textrm{edit\_distance}(t_j, \textrm{meeting } i)$}  
  \ENDFOR \label{alg:cal_edit_dist2}
  \STATE{$d_i^{\tprime}\leftarrow \displaystyle\min_j (d_{ij}^{\tprime})$} \label{alg:min_edit_dist}
  \ENDFOR
  \STATE{$F(\tprime) \hspace*{-2pt}\leftarrow \hspace*{-2pt} \frac{1}{m}\displaystyle\sum_{i} d_i^{\tprime} +C_1\textrm{length}(\tprime) $+$C_2\textrm{backw}(\tprime)$} \label{alg:costfun}
  \IF{ $F(\tprime) < F(t)$ } \label{alg:accept_jump1}
  \STATE{$t \leftarrow \tprime$, accept jump with probability 1, $\iter\leftarrow \iter+1$, $F(t)\leftarrow F(t')$}
  \IF {$F(\tprime) < F(t_{\textrm{best}})$}
  \STATE{$t_{\textrm{best}}\leftarrow \tprime $}
  \ENDIF
  \ELSE
  \STATE{$t \leftarrow \tprime$ with probability  $\exp \left(  -\Delta F/T \right)$ where}
  \STATE{$\Delta F = F(\tprime)- F(t) $. If accepted, then $\iter\leftarrow \iter+1$, $F(t)\leftarrow F(t')$}  
  \ENDIF \label{alg:accept_jump2}
  \STATE{$T \leftarrow T_0* 0.95^{\iter}$ (update temperature)}
  \ENDWHILE
\end{algorithmic}
\end{algorithm}

\begin{figure}
\begin{center}
\includegraphics[width=3.2in]{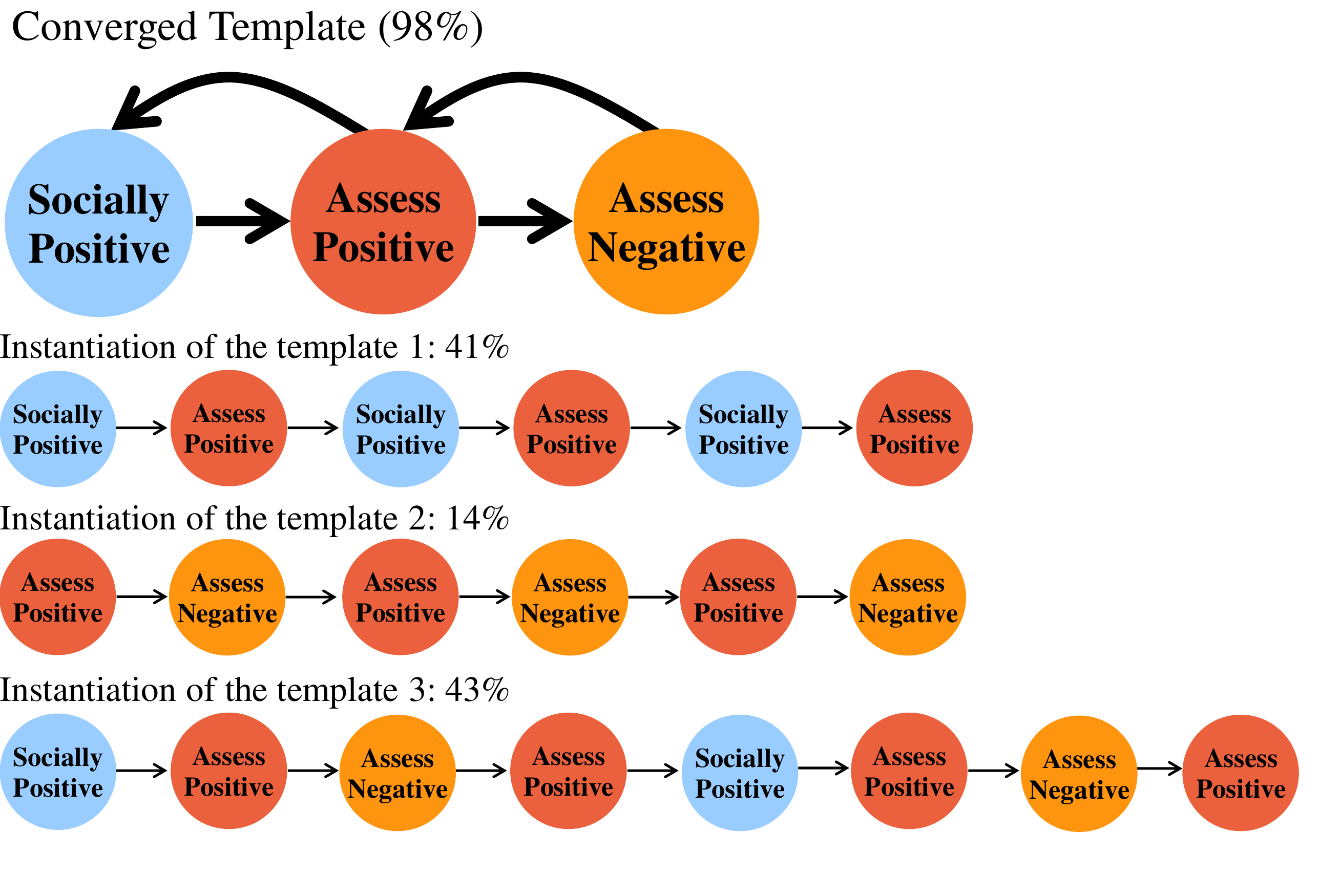}
 \caption{\label{quit_result} The template representing the interaction of social and work related acts. Specific template instantiations and how often they occurred in the experiment are also provided.}
\end{center}
\end{figure}

We ran the algorithm starting from 95 different initial conditions, each initial condition corresponding 
to one of the true meetings in the database. These 95 experiments were performed in order to find a fairly
 full set of reasonable templates, and to also determine whether the algorithm consistently settled on a
 small set of templates. The results were highly self-consistent, in the sense that in 98\% of the total 
experimental runs, the algorithm converged to a template that is equivalent to, or a simple instantiation of, the one shown in 
Figure \ref{quit_result}. This template has a very simple and strong message, which is that the next judgment following a negative assessment is almost never a socially positive act. The converse is also true, that a socially positive act is rarely followed by a negative assessment.

To assess the correctness 
of this template, we note that of the 1475 times a socially positive act appears, it is adjacent to a 
negative assessment 141 times. Of the 991 times a negative assessment appears, 132 times a socially positive act 
is adjacent to it. One can contrast these numbers with the
percent of time a positive assessment is adjacent to a socially positive act (68\%), though there are generally more positive 
assessments than either socially positive acts or negative assessments, which intrinsically lowers the probability that a socially positive act would be adjacent to a negative assessment; however, even knowing this, there could be more powerful reasons why socially positive acts are associated with positive assessments rather than negative assessments.

From a social perspective, this result can be viewed as somewhat counterintuitive, as one might imagine wanting 
to encourage a colleague socially in order to compensate for a negative assessment. In practice, however, sometimes 
positive social acts can sound disingenuous when accompanying a negative assessment. This can be demonstrated using
 the following pairs of dialog from within the AMI Corpus (that are not truly side by side): 1) ``But I thought it
 was just completely pointless.'' ``Superb sketch by the way.'' 2) ``Brilliantly done.'' ``It's gonna inevitably sort 
of start looking like those group of sort of ugly ones that we saw stacked up.'' 3) ``It'd be annoying.'' 
``Yeah, it was a pleasure working with you.'' 4) ``No, a wheel is better.'' ``All thanks to Iain for the design of that one.''
In these cases a positive social act essentially becomes a negative social act.

The algorithm introduced here found a macro pattern of assessments that is prevalent and yields insight into
 the interaction between social and work related dialogue acts. This demonstrates that machine learning can be used 
to illuminate fundamental aspects of this domain. 
This knowledge is general, and could be used, for instance, to help classify assessments when they are not labeled,
 and to potentially help reconstruct missing/garbled pieces of dialogue. These are necessary elements in building any type of meetings analysis assistant. To our knowledge, the macro pattern we found has not been previously noted in the literature.

\subsection{Experimental Comparison with Other Methods}

In this section, we compare our results with other methods, particularly profile Hidden Markov Model (HMM) and Markov chain.
We focus on comparison of results in this section, however, further in depth discussion is provided in Section \ref{relatedwork_qii}.

Profile HMM is a tool to find and align related sequences, and match a new sequence to known sequences. The structure of a profile HMM has forward arrows between match states, along with delete and insertion nodes that allow any string to fit into the template. Profile HMM shares similar features with our algorithm, in that it finds a common sequence given sequences with different lengths. To create a profile HMM, there are a series of heuristic steps. One way of learning profile HMM, used to produce Table \ref{tab:profileHMM}, is the following. First, we choose the length of the HMM. Second, we estimate the parameters in the HMM model using pseudo-counts for the prior. Third, we find the most likely string to be emitted from each match state.

As mentioned above, the length of a profile HMM is generally specified by the user. Table \ref{tab:profileHMM} shows profile HMMs with a variety of lengths specified. We were not able to recover the full pattern with profile HMM that our method produced, but we were able to see, for instance, that socially positive acts are often next to positive assessments, which is one of the instantiations of the template our method discovered.


\begin{table}
\caption{Profile HMM (SP: Socially positive, AP: Assess positive, AN: Assess negative)\label{tab:profileHMM}}
\begin{center}
    \begin{tabular}{|p{2cm}| p{9.6cm}|}
    \hline
\bf{Length specified} & \bf{Result} \\ \hline
3&SP $\rightarrow$ AP $\rightarrow$ AP\\
5&SP $\rightarrow$ AP $\rightarrow$ SP $\rightarrow$ AP $\rightarrow$ AP\\
10&SP $\rightarrow$ AP $\rightarrow$ SP $\rightarrow$ AP $\rightarrow$ AP $\rightarrow$ AP $\rightarrow$ AP $\rightarrow$ AP $\rightarrow$ AP $\rightarrow$ AP\\
20&SP $\rightarrow$ AP $\rightarrow$ SP $\rightarrow$ AP $\rightarrow$ AP $\rightarrow$ AP $\rightarrow$ AP $\rightarrow$ AP $\rightarrow$ AP $\rightarrow$ AP $\rightarrow$ AP $\rightarrow$ AP $\rightarrow$ AP $\rightarrow$ AP $\rightarrow$ AP $\rightarrow$ AP $\rightarrow$ AP $\rightarrow$ AP $\rightarrow$ AP $\rightarrow$ AP\\
\hline
    \end{tabular}
\end{center}
\end{table}

\begin{figure}[h!]	
\begin{center}
\includegraphics[width=2.5in]{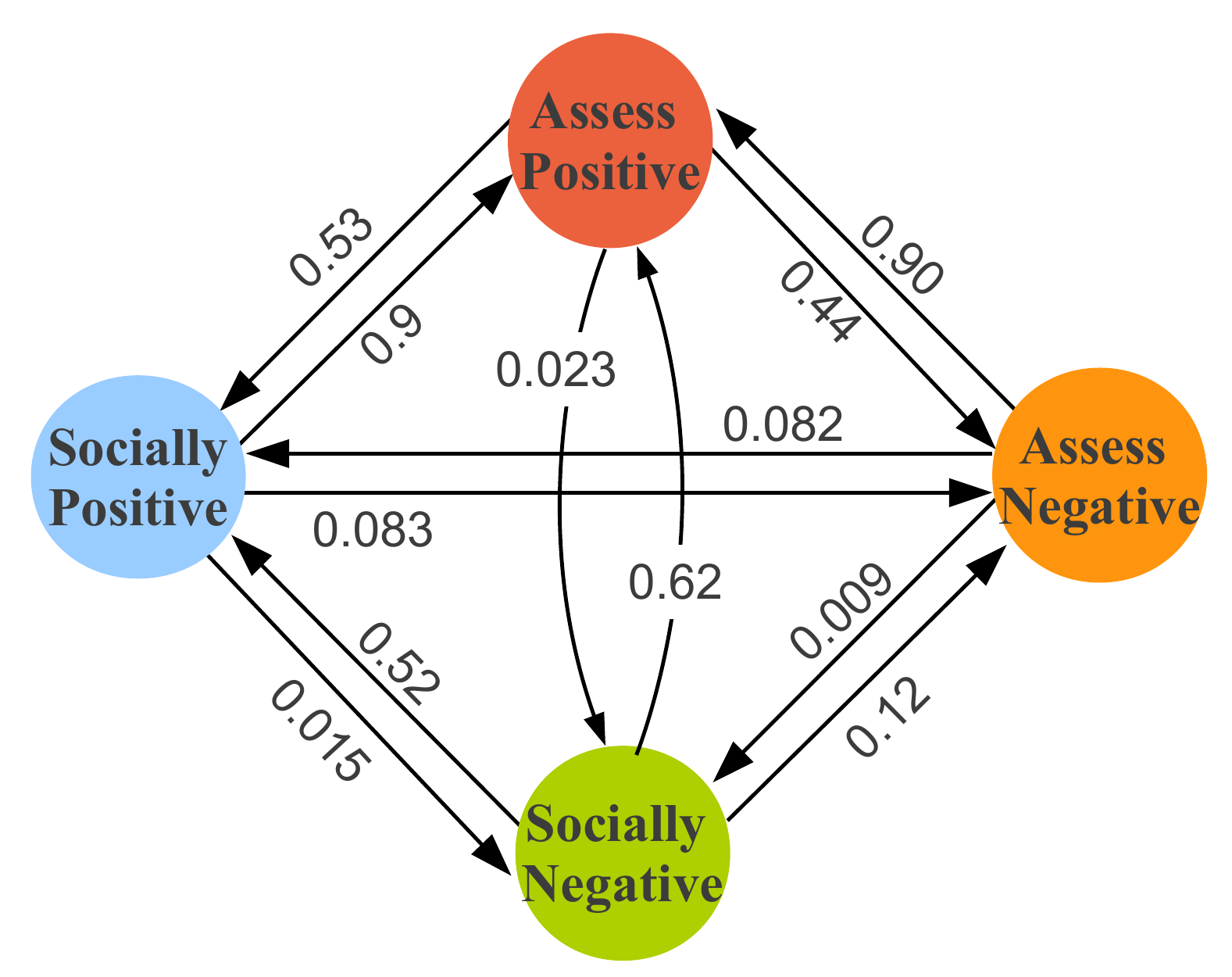}
 \caption{\label{fig:markov} Learned Markov Chain}
\end{center}
\end{figure}

We note that a first-order Markov chain could be used to model transitions between nodes, however, since a Markov chain records all possible transition probabilities, it is not clear in general how to turn this into a template that would yield insight. On the other hand, once we provide a template like the one in Figure \ref{quit_result}, it is easy to see the template within the Markov chain. The first-order Markov chain for the meetings data, learned using maximum likelihood estimation, is provided in Figure \ref{fig:markov}. This learned Markov chain supports our 
learned template, as the four highest transition probabilities exactly represent our learned template --- with high transition probabilities between socially positive and assess positive, and between 
assess negative and assess positive. The reason that socially negative did not appear in Figure \ref{quit_result} (as it should not) is because the number of socially negative dialogue acts is much smaller than the number of other dialogue acts (only 1\%). This is an important difference between a Markov chain and our algorithm --- our algorithm picks the dominant macro pattern, and has the ability to exclude a state (i.e. dialogue act) if there is no strong pattern involving that state. The Markov chain, on the other hand, is required to model transition probabilities between all states that exist within the data. We also note that it is possible for our method to choose forward arrows with very low transition probabilities in the Markov chain, so looking at only the high probability transitions will not suffice for constructing a template that minimizes our risk functional.


\section{Question iii: How long is this meeting going to last given 
that the decision has already been made?\label{q3}} 
Meetings sometimes last longer than expected. Even when all the decisions 
seem to be made, it often takes some time to work out the details and formally end the meeting.
We would like to know whether it is possible to predict when the meeting is 
going to be over if we know when the decisions are made. 

\begin{figure}
\begin{center}
\subfigure[Total meeting time (red) and discussion ending time (blue).]{ 
  \includegraphics[width=4.5in]{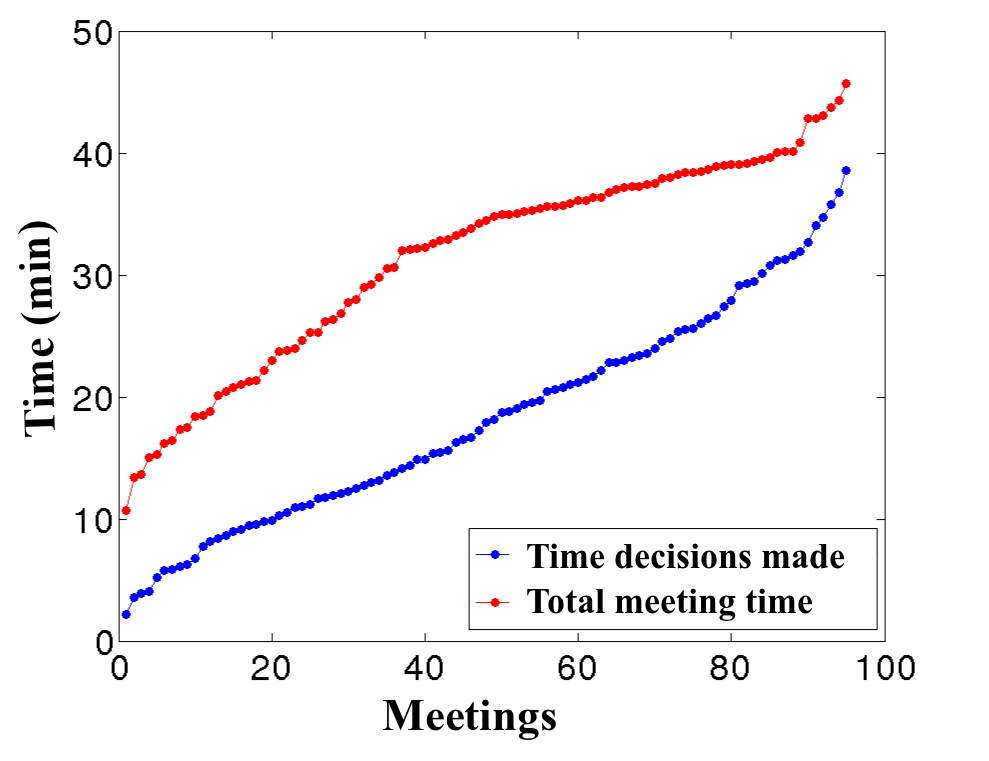}
}
\subfigure[``Wrap-up'' time vs$.$ time until decisions are made.
]{ 
  \includegraphics[width=4.5in]{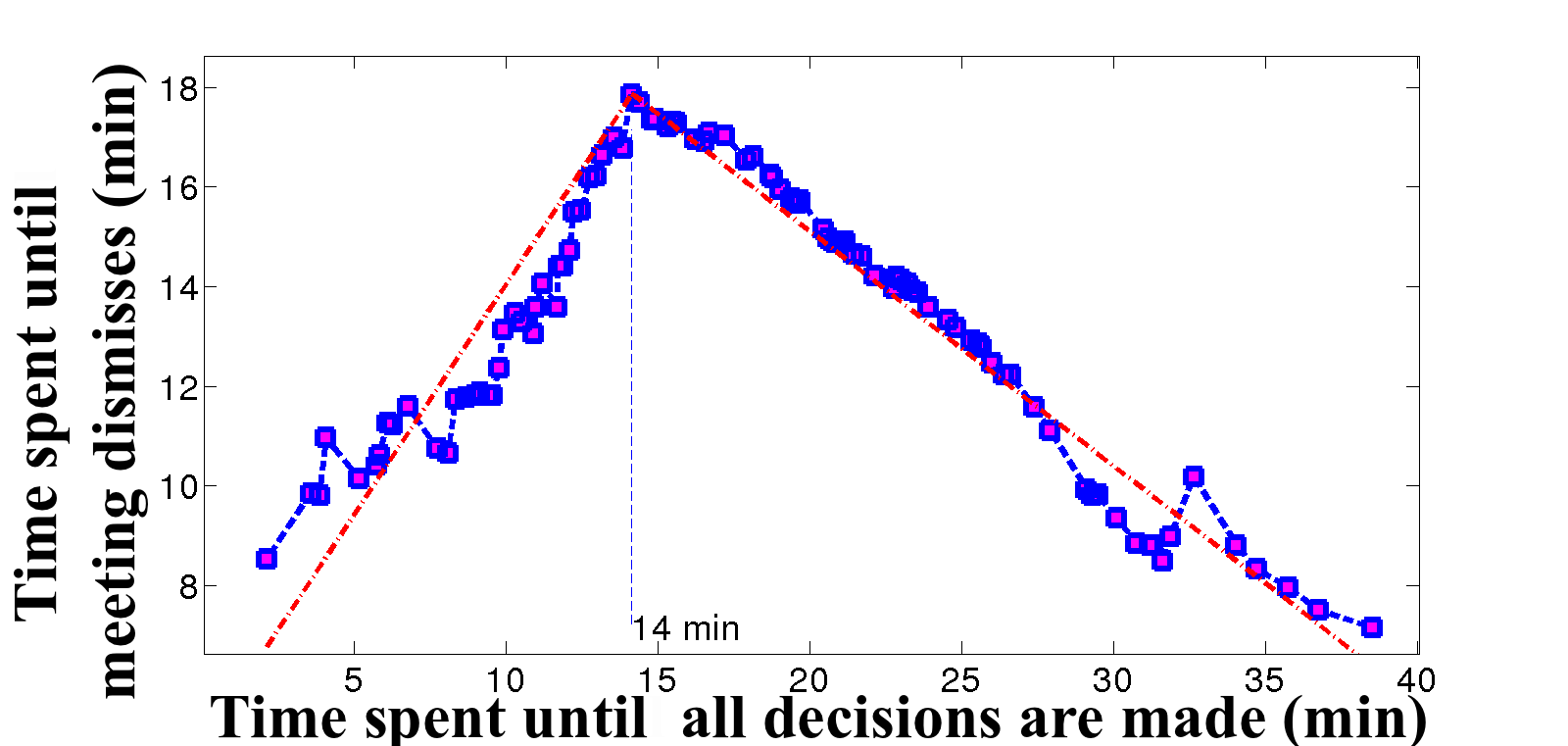}
}
\caption{Total meeting time and wrap-up time\label{discussion_patch}}
\end{center}
\end{figure}

Figure \ref{discussion_patch}(a) displays meetings along the horizontal axis, ordered by total meeting time. The annotated time when key decisions are all made (discussion finishing times) are indicated by blue squares, and the total meeting times are indicated by red squares for each meeting.
 Figure  \ref{discussion_patch}(b) shows the time between when key decisions are made on the x-axis, 
and the wrap-up time on the y-axis (this is the time spent after key decisions are made and before the meeting finishes).
With a simple piecewise linear formula, one can predict with relatively high accuracy what the wrap-up time will be, given the time at which the key decisions are finished being made. Denoting the time to complete key decisions by $x$, the estimated wrap-up time is as follows:
If $x\leq$  about 14 minutes, the wrap-up time is about $0.923  + 4.78$ minutes.
 If $x>$ about 14 minutes, the wrap-up time is $-0.47x + 24.53$ minutes.
There are some interesting implications: if the meeting was efficient enough to make all the decisions within 14 minutes, 
then the team will also be efficient in wrapping up. Once the meeting has gone about 14 minutes without all decisions made, then people also tend to spend more time wrapping up. If the meetings runs very long without decisions being made, then once the decisions are made, the meeting tends to wrap up quickly.

These results are specific to meetings whose length is less than an hour as in the AMI corpus, 
which, according to \citet{romano2001meeting}
is true 26\% of the time; it would be interesting for future work to see 
if a piecewise linear model works well for meetings of other lengths, though this hypothesis cannot currently be verified by any database that we are aware of.

\section{Question iv: Do persuasive words exist?\label{q4}}
A ``good'' meeting might be one where we have contributed to the team effort. We always want to suggest good ideas and want the team members to accept those ideas. 
Numerous articles claim that how we package our ideas, and our choice of words, is sometimes as important as the idea itself \citep{persuasivewords1,persuasivewords2}. We are interested in understanding this hidden factor in the team's decision making process. Are there patterns in suggestions that are accepted versus rejected? Can we use this to improve how we present ideas to the team? 

To select data for this section, we chose a bag-of-words representation for each ``suggestion'' dialogue act.  We gathered a set of all words that occurred in all suggestions, excluding stop words, leading to a 1,839 dimensional binary vector representation for each suggestion. The labels were determined by the annotations, where accepted suggestions received a $+1$ label, and rejected suggestions received a $-1$ label.

\subsection{Are published sets of persuasive words really persuasive?\label{sec:empiricalPersuasive}}
We would like to know whether the persuasive words of \citet{persuasivewords1,persuasivewords2} are truly 
persuasive. These words are not domain-specific, meaning that they do not depend on the topic of the meeting, and could thus be more generally useful.
Note that the data cannot tell us (without a controlled experiment) whether there is a truly causal 
relationship between using persuasive words and having a proposal accepted; however, we can study correlations, 
which provide evidence. 
An important first question is whether the proportion of persuasive words in accepted suggestions differ significantly
from the proportion of persuasive words in rejected suggestions. We marked each suggestion
 as to whether or not one of the words in  \citet{persuasivewords1,persuasivewords2} appears. We also
 mark each suggestion as to whether it is an accepted or rejected suggestion. Of the 139 times that 
a suggestion contained persuasive words from \citet{persuasivewords1,persuasivewords2}, 134 of these 
appearances were within accepted suggestions (96\% of appearances). Of the 2,185 times a suggestion did 
not contain any of the words from \citet{persuasivewords1,persuasivewords2}, 
1,981 of these appearances were within accepted suggestions (90\% of appearances). Using Fisher's exact test, 
the difference in proportions is significant at the 0.01 level (pvalue 0.0037); it appears that persuasive words do appear 
more often in accepted suggestions than other words do. 

On the other hand, we tested each persuasive word individually, to see whether the number of accepted suggestions
 it appears in is significantly different than the proportion of rejected suggestions it appears in. The difference
 between the two proportions was not significant at the 0.05 level for any of the persuasive words, again using Fisher's
 exact test for differences between proportions. Perhaps this is because each word individually is rare, while the
 collection is not. This begs the question as to whether there are words that themselves show a significant difference 
in the proportion of accepted suggestions they appear in. A related question is which words are the most important,
 in terms of predictive modeling, for distinguishing accepted and rejected suggestions. We will answer both of these questions below, the latter answered first.

\nopagebreak[4]

\subsection{Using SVM coefficients to tell us about persuasive words\label{svm_feature_ranking}}
In this section, we try to discover persuasive words from data alone, and afterwards compare to the published lists of persuasive words.
Specifically, we want to know whether we can predict if a suggestion would be accepted or rejected solely based on words that are used,
 not based on the idea within it. A standard way to do this is to apply a linear SVM with cross-validation for the tradeoff parameter, and examine the
 values of its coefficients; this provides a ranking of features, and a rough measure of how important each word is to the overall 
classifier \citep[e.g., see][]{Guyon02}.
The set of 2,324 suggestions with 1,839 features (one feature per unique word) was separated into 5 folds using each in turn as the test set, and the SVM accuracy was 83\% $\pm$ 2.1\%.
Note that the SVM has no prior knowledge of what persuasive words are. Persuasive words identified by the SVM are considered to be words with large absolute coefficient value (normalized) over all 5 folds. These include ``things,'' ``start,'' ``meeting,'' ``people,'' ``yeah.'' Many of the persuasive words concerned the topic of marketing (``market,'' ``presentation,'' ``gimmick,'' ``logo'') and non-persuasive words included ``buttons,'' ``speech,'' ``LCD,'' ``recognition,'' and more often words specific to the topic of creating a remote control (``scroll,'' ``green'').

The next question is whether these learned persuasive words make sense. To answer this, we checked whether the persuasive words from \citet{persuasivewords1,persuasivewords2} had positive or negative SVM coefficients. The overlap between our set of 1,839 words and the persuasive words from 
\citet{persuasivewords1,persuasivewords2} is 29 words. Their average coefficient values over 5 folds
are shown in Table \ref{tab:listofpersWordsCompare}. The finding is consistent with the Fisher's exact
test result in Section \ref{sec:empiricalPersuasive} --- not a single word has a large positive coefficient and 
all words have relatively large standard deviations, showing 
lack of their individual connection to persuasiveness. 

\begin{table}
\caption{List of persuasive words on which SVM and the articles \citep{persuasivewords1,persuasivewords2} agree/disagree
\label{tab:listofpersWordsCompare}}
\begin{center}
    \begin{tabular}{|  p{3cm} | p{3cm} | p{3cm} | p{3cm} |}
    \hline
\multicolumn{2}{|c} { \bf{ Agrees }}  &  \multicolumn{2}{|c|} { \bf{Disagrees} } \\ \hline
\bf{ Words  }&  \bf{$\lambda$ $\pm$ Std.}                   &    \bf{Words}  & \bf{$\lambda$ $\pm$ Std.} \\ \hline  
  strength	&	0.006	$\pm$	0.0026	  &    inspiration	&	-0.0002	$\pm$	0.003	\\
    free	&	0.0054	$\pm$	0.0031	  &    drive	&	-0.0002	$\pm$	0.0011	\\
    good	&	0.0048	$\pm$	0.0015	  &    easy	&	-0.0004	$\pm$	0.003	\\
    power	&	0.0043	$\pm$	0.0032	  &    health	&	-0.0005	$\pm$	0.0022	\\
    avoid	&	0.0039	$\pm$	0.0011	  &    creativity	&	-0.0016	$\pm$	0.0021	\\
    offer	&	0.0036	$\pm$	0.0035	  &    guarantee	&	-0.0017	$\pm$	0.001	\\
    save	&	0.003	$\pm$	0.0036	  &    explore	&	-0.0017	$\pm$	0.001	\\
    safe	&	0.0028	$\pm$	0.0015	  &   safety	&	-0.0017	$\pm$	0.001	\\
    energy	&	0.0024	$\pm$	0.0032	  &    reinvent	&	-0.0017	$\pm$	0.0009	\\
    imagine	&	0.0024	$\pm$	0.0016	  &    approach	&	-0.0019	$\pm$	0.003	\\
    important	&	0.0023	$\pm$	0.002	  &    money	&	-0.0025	$\pm$	0.0035	\\
    confidence	&	0.0019	$\pm$	0.0027	  &    purpose	&	-0.0027	$\pm$	0.0008	\\  
    wanted	&	0.0015	$\pm$	0.0009	 &  & \\
    quick	&	0.0015	$\pm$	0.0009	 &  & \\
    memory	&	0.0015	$\pm$	0.0009	 &  & \\
    life	&	0.0006	$\pm$	0.0029	 &  & \\
    hurry	&	0.0005	$\pm$	0.0029	 &  & \\ \hline
    \end{tabular}
\end{center}
\end{table}

In what follows we will explore a different method for feature ranking, where the space is reduced 
to only words that occur in significantly different proportions between accepted and rejected suggestions before applying an SVM.

\subsection{Which words are individually persuasive?}
We have 2,324 suggestions in an 1,839 dimensional space. Fisher's exact test is based on the hypergeometric distribution and has the benefit that it can be used for features that are rarely 
present. It considers whether the suggestions containing the feature have significantly different proportions of acceptance and rejection. The selected persuasive words are shown in 
Table \ref{tab:persWordsFishers} with their associated pvalues, where the most significant words are at the top of the list. We can also find non-persuasive words that are significant 
according to Fisher's exact test. Together with the persuasive words, this gives us a set of important features to use as a reduced feature space for machine learning. Some of the 
non-persuasive words include ``recognition'', ``speech'', ``fair'', ``selecting'', ``flat'', ``animals'', ``middle'', and ``bottom''. The total number of features we used 
is 244 (where persuasive words and non-persuasive words were selected under the same significance threshold). After this feature reduction, we applied SVM and achieved an accuracy 
of 87.2\% $\pm$ 0.010\%, which is 
higher than before. 
The SVM coefficients for all of the words that we identified as being persuasive are positive, as shown also in Table \ref{tab:persWordsFishers}. 
These words are thus persuasive both individually and together: they are each individually significant in predicting accepted suggestions, and they have positive predictive coefficients.

\begin{table}
\caption{List of persuasive words from Fisher's exact test (higher ranking for smaller pvalues). The second column contains the pvalue
 from Fisher's exact test. The third column contains the ratio of accepted proposals when the word appeared. The fourth column 
contains the ratio of accepted proposals when the word did not appear. The last column contains the SVM coefficients.\label{tab:persWordsFishers}}
\begin{center}
    \begin{tabular}{| p{1.8cm} | p{1.4cm} | p{3cm} | p{3cm} | p{2.6cm}| }
    \hline
\bf{Words}& 
 \bf{Pvalues}&  
\bf{Ratio of accepted when appears}&
\bf{Ratio of accepted when not appears}& 
\bf{SVM $\lambda$ $\pm$ Std.} \\ \hline
yeah &   0.012 & 1 (46/46) & 0.90 (2069/2278) &  0.011 $\pm$ 0.0021\\ 
give &   0.020 & 1 (41/41) & 0.90 (2074/2283) &  0.010 $\pm$ 0.0015 \\ 
menu &   0.027 & 1 (38/38) & 0.90 (2077/2286)  &0.010 $\pm$ 0.0045\\ 
start &   0.043 & 1 (33/33) & 0.90 (2082/2291)  &0.012 $\pm$ 0.0020 \\ 
meeting &   0.052 & 1 (31/31) & 0.90 (2084/2293) & 0.012 $\pm$ 0.0019\\ 
touch &   0.085 & 1 (26/26) & 0.90 (2089/2298)   & 0.0081 $\pm$ 0.00059\\ 
discuss &   0.093 & 1 (25/25) & 0.90 (2090/2299)  & 0.0092 $\pm$ 0.0024\\ 
find &   0.093 & 1 (25/25) & 0.90 (2090/2299)  & 0.0098 $\pm$ 0.0017\\ 
market &   0.12 & 1 (22/22) & 0.90 (2093/2302)  & 0.0094 $\pm$ 0.0023\\ 
yellow &   0.12 & 1 (22/22) & 0.90 (2093/2302)  & 0.0076 $\pm$ 0.0018\\ 
work &   0.12 & 1 (22/22) & 0.90 (2093/2302)  & 0.0097 $\pm$ 0.0031\\ 
good &   0.13 & 0.97 (37/38) & 0.90 (2078/2286)  &0.0029 $\pm$ 0.0035\\ 
fruit &   0.13 & 1 (21/21) & 0.90 (2094/2303)  & 0.0065 $\pm$ 0.0032\\ 
logo &   0.15 & 1 (20/20) & 0.90 (2095/2304)  & 0.0073 $\pm$ 0.0019\\ 
people &   0.15 & 0.97 (35/36) & 0.90 (2080/2288)  & 0.012 $\pm$ 0.0024\\ 
side &   0.16 & 0.97 (34/35) & 0.90 (2081/2289)  & 0.0056 $\pm$ 0.0039\\ 
number &   0.16 & 1 (19/19) & 0.90 (2096/2305)  & 0.0059 $\pm$ 0.0027\\ 
presentation &   0.18 & 1 (18/18) & 0.90 (2097/2306)  &0.0084 $\pm$ 0.0020\\ 
things &   0.19 & 0.95 (45/47) & 0.90 (2070/2277)  & 0.012 $\pm$ 0.0026\\ 
chip &   0.20 & 1 (17/17) & 0.90 (2098/2307)  & 0.0066 $\pm$ 0.0026\\ 
stick &   0.22 & 1 (16/16) & 0.90 (2099/2308)  & 0.0070 $\pm$ 0.0017\\ 
gonna &   0.22 & 0.95 (42/44) & 0.90 (2073/2280)  & 0.000063 $\pm$ 0.0057\\ 
information &   0.24 & 1 (15/15) & 0.90 (2100/2309)  & 0.0059 $\pm$ 0.0020\\ 
talk &   0.24 & 1 (15/15) & 0.90 (2100/2309)  &0.0034 $\pm$ 0.0027\\ 
\hline
\end{tabular}
\end{center}\end{table}

Studying the most persuasive words from Fisher's exact test, we find that many of these words are \textit{not} specifically tied to the topic of the meeting (designing a remote control), but seem to be more generally persuasive. In fact, we can make informed hypotheses about \textit{why} these words are persuasive. Some of these observations help us to understand more generally how language is used during meetings.
Let us consider the most significant persuasive words as follows:
\begin{itemize}
\renewcommand{\labelitemi}{$\bullet$}
\item \textit{Yeah:} Dialogue segments where the word ``yeah'' is used include: ``or yeah, maybe even just a limited multi-colour so
 it doesn't look too childish,'' ``yeah, if you had one of those, just coming back to your other point about pressing the button and 
setting off the bleeper in the room,'' ``Yeah if you are holding it in your hand you could do that.'' Judging from these and similar 
dialogue segments, our hypothesis is that framing a suggestion as an agreement with a previous suggestion increases its chances of being 
accepted. That is, if the idea comes across as if it were in line with previous thoughts by others, the suggestion has a higher chance of being accepted.
This applies either when attributing the full idea to others, or just the line of thought. The case where one attributes their full idea to 
others in order to increase its chances of acceptance has been considered in popular books \cite{carnegie2010win}.

\item \textit{Give:} ``Give'' is used in at least three ways, the first one occurring the most often: (i) giving with respect to the topic 
of the meeting, which here is either the customer or the product (``so if you want to give the full freedom to the user,'' ``You give it the 
full functions in here,'' ``We can give them smooth keys''), (ii) giving to the meeting participants (``would give us a little bit of a 
marketing niche''), and (iii) to indicate that suggestions are based on previous data or knowledge (``given these parameters that we're just 
gonna sort of have this kind of uh non-remote remote'', ``given speech recognition I think you should go for the less fancy chip'').

\item \textit{Menu:} This word seems to be tied to the topic of remote controls (``Um and one other suggestions I'd make is to in is to include in 
a menu system''), without a general hypothesis for other types of meetings.

\item \textit{Start:} Our hypothesis is that the word ``start'' gives group members the opportunity to agree, where agreement of basic suggestions
 provides an indication that group members want to be productive during the meeting; e.g., ``Shouldn't we start with the most important parts?'' 
``I will start by the basic one.'' This type of agreement may help with alliance building early on in the meeting.

\item \textit{Meeting:} The word ``meeting'' often appears in suggestions about what \textit{not} to discuss: ``Or maybe this is something for the 
next meeting,'' ``I figure we could get back to it on the next meeting actually,'' ``We take it to the other meeting.'' Our hypothesis is that 
suggesting that a topic belong to a later meeting may be a way to gently change the topic or move the current meeting along. It can be used instead of 
a negative assessment of a previous suggestion.

\item \textit{Touch:} This seems to be tied to the topic of remote controls, ``So we put a touch pad on it,''  ``We can uh do a touch-pad on our remote.''

\item \textit{Discuss:} This word appears mainly in an organizational context for the meeting: ``And then we can discuss some more closely,'' ``I 
think we shouldn't discuss any points points that long,'' ``Maybe we should centralise the discussion here.'' It seems that often, people tend to agree
 with organizational suggestions about the meeting.

\item \textit{Find:} Our hypothesis is that ``find'' is often used in suggestions to gather more information or do more work, and these are suggestions 
that are often accepted:  ``we have to find out if it's possible,'' ``um and I'm sure we can find more goals for the product we are going to develop,'' 
``but just try to find out what they're willing to pay for it.''

\end{itemize}

Having a collection of persuasive words can be immediately useful, assuming that there is a causal relationship between persuasive words and accepted proposals, rather than only a correlation. The hypothesis tested in this work  (that persuasive words truly exist) can also be tested for this causal relationship, and it would be very interesting to  create a dataset for this purpose. If the hypothesis does hold, it has the potential to allow ideas to be communicated more clearly, and thus to make meetings more efficient overall. Already we have gained some insight for how specific words are used within meetings, and why suggestions containing these words are more likely to be accepted.
 
\section{Related Work\label{related_work}}

The closest work to ours is that of the CALO (Cognitive Assistant that Learns and Organizes) project \citep{calo}, which 
takes a \textit{bottom-up} approach to designing a cognitive assistant that can reason and learn from users
using machine learning techniques. Although the CALO project's focus (e.g., speech recognition, sentence segmentation, dialogue act tagging, 
topic segmentation, action item detection and decision extraction) does not intersect
with our work, it is worthwhile to note that this multi-year project has 
made a definite step towards improving meetings using machine learning techniques. 
However, as pointed out by \citet{tur2008calo}, a number of challenges 
(e.g.\ extracting task descriptions) still exist
before this tool can be integrated into daily life.
The insights into meetings obtained through our work could be used as part of a 
\textit{top-down} approach to design such a tool.

We believe our work is the first to take a truly data-driven approach to finding persuasive words; this is the first work to try to prove that a word is persuasive using data.


\paragraph{Qualitative Work:}
Research on group decision making processes traditionally
appears in \textit{qualitative} studies. Works along these lines attempt to 
understand and model the course of agreement \citep{Black48}, 
design features to track the course of negotiation \citep{Eugenio99theagreement},
develop theories for group decision making \citep{davis1973group}
and study different group decision-making schemes \citep{green1980effects}.

\paragraph{Related Work for Question i:} 
A few quantitative approaches have attempted to detect when decisions are made. These methods use
maximum entropy \citep{hsueh2008automatic}, linear SVMs \citep{fernandez2008modelling}
and directed graphical models \citep{bui2009extracting}. 
A common factor of these works is that they require content information (audio and video data; prosodic features, position within the meeting, length in words and duration etc.) that can potentially contain sensitive information about what is being planned.
Our approach requires only dialogue acts (i.e.\ the actual meeting content is not required) to achieve the same goal with relatively high accuracy. One benefit of using only dialogue acts is that 
the algorithm allows our results to hold independently of the specific type 
and purpose of the meeting being held and can be used in situations 
where the meeting contains potentially sensitive information. It is also much simpler logistically to require collecting only dialogue acts.

\paragraph{Related Work for Question ii: \label{relatedwork_qii}}

To the best of our knowledge, our work is the first to apply learning techniques to learn dynamical interactions
between social and work aspects in meetings. As discussed earlier, one could consider the method we developed generally for
learning a template from a sequence of data. Our method has the following characteristics: 1) The loss function
considers meetings instantiated from the same template, but possibly with different lengths, to be equally good.
For instance: if a template is `A' `B' `C' with a back arrow from `C' to `A', then `ABCABC', `BCABCA', `CAB' are equally good
instantiations from this template. 2) An instantiation of the template can start at any position, as demonstrated by
the `BCABCA' and `CAB' instances in the template above. 3) By optimizing the loss function we provide in Section \ref{q2},
the algorithm is able to uncover a template of exactly the desired form, despite the existence of noise. Let us compare
this method to other graph learning techniques within related subfields.

\textit{Petri nets} are used (and were designed) to model concurrency and synchronization in distributed systems. In
learning Petri nets from data, and generally in learning workflow graphs and workflow nets (see \cite{van1998application}), all
possible transitions in the data must be accounted for within the learned net (e.g., \cite{agrawal1998mining}), which is the opposite
of what we would like to do, which is to create instead a more concise representation that is not necessarily all-inclusive.
The kind of templates we discover should be much simpler than Petri nets - for us, there is only sequential and iterative
routing, and no parallel routing (no concurrency), there are no ``tokens'' and thus there is no underlying workflow state, and
there are no conditions involved in moving from one place to the next (no conditional routing).

One major difference between our approach to learning macro-patterns and learning probabilistic graphical
models such as Markov models is that our approach is deterministic (not probabilistic). We do not model the
probabilities of transitions between states, as these probabilities are not of fundamental interest for building the template (but
could be calculated afterwards). As in other deterministic methods (e.g., SVM), we aim to directly optimize
the quality measure that the macro-pattern will be judged by on the data, in our case involving edit distance.
This allows us to handle noisy data, that is, meetings that do not fit precisely into the pattern, without having
to include additional nodes that complicate the template. We do not then need a graph that handles all possible
transitions and their probabilities (as is required in Markov models or Petri nets). It is certainly possible to
model a macro-pattern as a Markov chain, as we demonstrated, but the resulting transition matrix would most likely
provide little insight. The full Markov transition matrix would be of size $|\Lambda| \times |\Lambda|$, but distilling
this to a compact one dimensional graph with mostly forward arrows would then be an additional task.

Hidden Markov Models aim to infer unobserved states, where in our case, there are no natural unobserved states.
Although we could artificially create hidden states, it is more natural to directly model the observed sequence. One exception to this is profile HMM's, where hidden states are either ``match'' states, ``insert'' states, or ``delete''
states \citep{eddy1998profile}. Methods for fitting profile HMM's generally do not create backwards edges, and thus cannot easily
accommodate substrings being repeated arbitrarily within the template. Profile HMM's are designed for accuracy in
aligning sequences, but they are not generally designed for conciseness or to minimize, for instance, a count of backwards
edges.
Methods for learning profile HMMs generally require the user to specify the length of the HMM, whereas we purposely do not do this.

Profile HMM's generally have a fixed initial state, whereas our method can start anywhere within the template.
The price of a fixed initial state and fixed length with no backwards edges is high; for instance for a template `A' `B' `C' with a
backwards arrow from `C' to `A,' we can equally well accommodate patterns `ABCABCABC' and `BCABCA,' as well as `BC', all
of them being perfect matches to our template. Profile HMM's, with any fixed initial state, and with no backwards arrows
to allow repeats, would require insertions and deletions for each of these patterns, and have difficulty viewing all of
these patterns equally. Note that in general, left-to-right HMM's cannot have backwards loops as our templates do.

One might also think of automata (\cite{narendra1974learning,wikiAutomata}) as a way to model meetings.
HMM's are equivalent to probabilistic automata with no final probabilities (see \cite{dupont2005links}), where above we discussed how
our goals do not generally involve hidden states. There are additional characteristics of automata that contrast with our
efforts. For instance, probabilistic automata generally have a set of initial states and accepting or final (terminating)
states, whereas our meetings can begin an end anywhere in the template, and there is no notion of acceptance or rejection
of a meeting to the template. Further, an important characteristic of Probabilistic Deterministic Finite Automata (PDFA) is
that for any given string, at most a single path generating the string exists in the automaton, whereas in our work, we do not require this uniqueness (\cite{guttman2006probabilistic}).

Probabilistic Suffix Automata (PSA) \citep{ron1996power} are probabilistic automata with variable memory length, that
aim to learn an order-$K$ Markov model with $K$ varying in different parts of state space. As with order-1 Markov chains
discussed above, is not clear how to construct a concise template of the form we are considering. On the other hand, it would
be an interesting extension of our work to keep track of higher order patterns within the template like PSA does.

One could think of our goal in Section \ref{q2} as solving a type of case of the consensus string problem \citep{sim2003consensus,lyngso2002consensus} with consensus error,
but with two major changes: (i) using edit distance between meetings and template instantiations as a metric, rather than between pairs of
strings, which allows backwards loops, (ii) encouraging conciseness in the template. Problems encountered when not including these aspects
were provided for the `ABC' example above.

The goal of learning metro maps \citep{shahaf2012trains} is very different than ours, since their goal is to learn a set of special paths
through a graph (metro lines) that have specific properties. Temporal LDA \citep{wang2012tm} is also very different than our work: its goal is to
predict the next topic within a stream, whereas our goal is to find a concise representation of a set of strings.

\paragraph{Related Work for Question iv:} 
Identifying characteristics of persuasive speech and discourse has been
of great interest in the qualitative research community
\citep{sternthal1978persuasive, scheidel1967persuasive}.
However, there has not been much quantitative research done on this topic. 
\citet{guerini2008trusting} studied the relationship between the choice of words
and the reaction they elicit in political speeches
based on numerical statistics, including counts of words in each document and how common the words are across all documents (tf-idf).
To the best of our knowledge, our work is the first work to apply machine learning techniques to learn persuasive words from 
free-form conversational meeting data, to rank persuasive words, and to compare resulting persuasive words with qualitative studies to gain insights.

\paragraph{Other meeting related work:}

Other studies that apply machine learning techniques for meeting related topics (but not specifically 
related to any of the work in this paper) include meeting summarization \citep{purver2007detecting},
topic segmentation \citep{galley2003discourse},
agreement/disagreement detection \citep{Hahn06, Hillard03detectionof,bousmalis2009spotting}, 
detection of the state of a meeting (discussion, presentation and briefing) \citep{banerjee2004using,reiter2005multimodal},
and prediction of the roles of meeting participants \citep{banerjee2004using}.
In all of this work, machine learning techniques are used as tools to 
classify a particular aspect of a meeting with a specific applications in mind. In addition to addressing different 
aspects of meetings, our work uses machine learning
techniques as a way to to learn and study meetings scientifically in an attempt to
bridge the gap between qualitative studies on understanding meetings and
quantitative application-focused studies.

\section{Conclusion\label{conclusion}}
The scientific field of meeting analysis is still in beginning stages; meetings have not yet been well characterized, and this is one of the first works in this new scientific arena.
 Several hypotheses made in this work \textit{cannot} be tested further with any current available dataset. We hope that by illuminating the potential of fully solving 
these problems, it will inspire the creation of new meetings corpora. 
Elaborating further, if we are able to automatically detect when key decisions are made,
 this could translate directly into a software tool that managers could use to determine when 
they should join an ongoing meeting of their staff without attending the full meeting. If we know common patterns of dialogue, 
this might help us to understand social cues better for business settings, and could potentially help reconstruct parts of 
dialogue that might not have been recorded properly. 
If we can automatically detect when a meeting's key decisions are made, 
and can accurately gauge the meeting wrap-up time, it can give us something immediately valuable, namely an estimated 
time for the end of the current meeting, which staff can use to plan ahead for the start of the next meeting, 
or to plan transportation, in a fast-paced corporate culture. 
If we truly knew which words were persuasive, we could use these words to 
help convey our ideas in the most favorable light. Before we can do all of this, however, we need to understand the science behind meetings, 
and make hypotheses that can be tested, which is the goal of this work.

\bibliography{final1}

\begin{thebibliography}{46}
\providecommand{\natexlab}[1]{#1}
\providecommand{\url}[1]{\texttt{#1}}
\expandafter\ifx\csname urlstyle\endcsname\relax
  \providecommand{\doi}[1]{doi: #1}\else
  \providecommand{\doi}{doi: \begingroup \urlstyle{rm}\Url}\fi

\bibitem[Agrawal et~al.(1998)Agrawal, Gunopulos, and
  Leymann]{agrawal1998mining}
R.~Agrawal, D.~Gunopulos, and F.~Leymann.
\newblock \emph{Mining process models from workflow logs}.
\newblock Springer, 1998.

\bibitem[Banerjee and Rudnicky(2004)]{banerjee2004using}
S.~Banerjee and A.I. Rudnicky.
\newblock Using simple speech based features to detect the state of a meeting
  and the roles of the meeting participants.
\newblock In \emph{Proceedings of International Conference on Spoken Language
  Processing}, pages 221--231, 2004.

\bibitem[Black(1948)]{Black48}
D.~Black.
\newblock On the rationale of group decision-making.
\newblock \emph{Journal of Political Economy}, 56\penalty0 (1):\penalty0
  23--34, 1948.

\bibitem[Bousmalis et~al.(2009)Bousmalis, Mehu, and
  Pantic]{bousmalis2009spotting}
K.~Bousmalis, M.~Mehu, and M.~Pantic.
\newblock Spotting agreement and disagreement: A survey of nonverbal
  audiovisual cues and tools.
\newblock In \emph{Proceedings of 3rd International Conference on Affective
  Computing and Intelligent Interaction and Workshops}, pages 1--9, 2009.

\bibitem[Bui et~al.(2009)Bui, Frampton, Dowding, and Peters]{bui2009extracting}
T.~H. Bui, M.~Frampton, J.~Dowding, and S.~Peters.
\newblock Extracting decisions from multi-party dialogue using directed
  graphical models and semantic similarity.
\newblock In \emph{Proceedings of the 10th Special Interest Group on Discourse
  and Dialogue Workshop on Discourse and Dialogue}, pages 235--243, 2009.

\bibitem[Carlson(2012)]{persuasivewords2}
G.~Carlson.
\newblock Uplifting word list:
  \url{http://www.freelance-copy-writing.com/Uplifting-Word-List.html}, 2012.

\bibitem[Carnegie(1936)]{carnegie2010win}
D.~Carnegie.
\newblock \emph{How to win friends and influence people}.
\newblock Simon \& Schuster, 1936.

\bibitem[Davis(1973)]{davis1973group}
J.~H. Davis.
\newblock Group decision and social interaction: A theory of social decision
  schemes.
\newblock \emph{Psychological Review}, 80\penalty0 (2):\penalty0 97--125, 1973.

\bibitem[Dupont et~al.(2005)Dupont, Denis, and Esposito]{dupont2005links}
P.~Dupont, F.~Denis, and Y.~Esposito.
\newblock Links between probabilistic automata and hidden markov models:
  probability distributions, learning models and induction algorithms.
\newblock \emph{Pattern Recognition}, 38\penalty0 (9):\penalty0 1349--1371,
  2005.

\bibitem[Eddy(1998)]{eddy1998profile}
S.~R. Eddy.
\newblock Profile hidden markov models.
\newblock \emph{Bioinformatics}, 14\penalty0 (9):\penalty0 755--763, 1998.

\bibitem[Ertekin and Rudin(2011)]{ErtekinRu11}
S.~Ertekin and C.~Rudin.
\newblock On equivalence relationships between classification and ranking
  algorithms.
\newblock \emph{Journal of Machine Learning Research}, 12:\penalty0 2905--2929,
  2011.

\bibitem[Eugenio et~al.(1999)Eugenio, Jordan, Thomason, and
  Moore]{Eugenio99theagreement}
B.~D. Eugenio, P.~W. Jordan, J.~H. Thomason, and J.~D. Moore.
\newblock The agreement process: An empirical investigation of human-human
  computer-mediated collaborative dialogues.
\newblock \emph{International Journal of Human Computer Studies}, 53\penalty0
  (6):\penalty0 1017--1076, 1999.

\bibitem[Fern{\'a}ndez et~al.(2008)Fern{\'a}ndez, Frampton, Ehlen, Purver, and
  Peters]{fernandez2008modelling}
R.~Fern{\'a}ndez, M.~Frampton, P.~Ehlen, M.~Purver, and S.~Peters.
\newblock Modeling and detecting decisions in multi-party dialogue.
\newblock In \emph{Proceedings of the 9th Special Interest Group on Discourse
  and Dialogue Workshop on Discourse and Dialogue}, pages 156--163, 2008.

\bibitem[Galley et~al.(2003)Galley, McKeown, Fosler-Lussier, and
  Jing]{galley2003discourse}
M.~Galley, K.~McKeown, E.~Fosler-Lussier, and H.~Jing.
\newblock Discourse segmentation of multi-party conversation.
\newblock In \emph{Proceedings of the 41st Annual Meeting of Association for
  Computational Linguistics}, pages 562--569, 2003.

\bibitem[Green and Taber(1980)]{green1980effects}
S.~G. Green and T.~D. Taber.
\newblock The effects of three social decision schemes on decision group
  process.
\newblock \emph{Organizational Behavior and Human Performance}, 25\penalty0
  (1):\penalty0 97--106, 1980.

\bibitem[Guerini et~al.(2008)Guerini, Strapparava, and
  Stock]{guerini2008trusting}
M.~Guerini, C.~Strapparava, and O.~Stock.
\newblock Trusting politicians' words (for persuasive {NLP}).
\newblock \emph{Computational Linguistics and Intelligent Text Processing},
  pages 263--274, 2008.

\bibitem[Guttman(2006)]{guttman2006probabilistic}
O.~Guttman.
\newblock \emph{Probabilistic Automata Distributions over Sequences}.
\newblock PhD thesis, The Australian National University, 2006.

\bibitem[Guyon et~al.(2002)Guyon, Weston, Barnhill, and Vapnik]{Guyon02}
I.~Guyon, J.~Weston, S.~Barnhill, and V.~Vapnik.
\newblock Gene selection for cancer classification using support vector
  machines.
\newblock \emph{Machine Learning}, 46\penalty0 (1-3):\penalty0 389--422, March
  2002.

\bibitem[Hahn et~al.(2006)Hahn, Ladner, and Ostendorf]{Hahn06}
S.~Hahn, R.~Ladner, and M.~Ostendorf.
\newblock Agreement/disagreement classification: Exploiting unlabeled data
  using contrast classifiers.
\newblock In \emph{Proceedings of the Human Language Technology Conference of
  the North American Chapter of the Association for Computational Linguistics},
  pages 53--56, 2006.

\bibitem[Hall(1994)]{hall1994americans}
J.~Hall.
\newblock Americans know how to be productive if managers will let them.
\newblock \emph{Organizational Dynamics}, 1994.

\bibitem[Hillard and Ostendorf(2003)]{Hillard03detectionof}
D.~Hillard and M.~Ostendorf.
\newblock Detection of agreement vs. disagreement in meetings: Training with
  unlabeled data.
\newblock In \emph{Proceedings of the Human Language Technology Conference of
  the North American Chapter of the Association for Computational Linguistics},
  2003.

\bibitem[Hsueh and Moore(2008)]{hsueh2008automatic}
P.~Y. Hsueh and J.~Moore.
\newblock Automatic decision detection in meeting speech.
\newblock \emph{Machine Learning for Multimodal Interaction}, pages 168--179,
  2008.

\bibitem[Ji and Bilmes(2005)]{ji2005dialog}
G.~Ji and J.~Bilmes.
\newblock Dialog act tagging using graphical models.
\newblock In \emph{Proceedings of International Conference on Acoustics,
  Speech, and Signal Processing}, volume~1, pages 33--36, 2005.

\bibitem[Kayser(1990)]{kayser1990mining}
T.~A. Kayser.
\newblock \emph{Mining group gold: How to cash in on the collaborative brain
  power of a group}.
\newblock Serif Publishing, 1990.

\bibitem[Lyngs{\o} and Pedersen(2002)]{lyngso2002consensus}
R.~B. Lyngs{\o} and C.~N.~S. Pedersen.
\newblock The consensus string problem and the complexity of comparing hidden
  markov models.
\newblock \emph{Journal of Computer and System Sciences}, 65\penalty0
  (3):\penalty0 545--569, 2002.

\bibitem[Mackenzie and Nickerson(2009)]{mackenzie2009time}
A.~Mackenzie and P.~Nickerson.
\newblock \emph{The time trap: The classic book on time management}.
\newblock Amacom Books, 2009.

\bibitem[McCowan et~al.(2005)McCowan, Carletta, Kraaij, Ashby, Bourban, Flynn,
  Guillemot, Hain, Kadlec, Karaiskos, et~al.]{Mccowan05theami}
I.~McCowan, J.~Carletta, W.~Kraaij, S.~Ashby, S.~Bourban, M.~Flynn,
  M.~Guillemot, T.~Hain, J.~Kadlec, V.~Karaiskos, et~al.
\newblock The {AMI} meeting corpus.
\newblock In \emph{Proceedings of Methods and Techniques in Behavioral
  Research}, 2005.

\bibitem[Nagata and Morimoto(1994)]{nagata1994first}
M.~Nagata and T.~Morimoto.
\newblock First steps towards statistical modeling of dialogue to predict the
  speech act type of the next utterance.
\newblock \emph{Speech Communication}, 15\penalty0 (3-4):\penalty0 193--203,
  1994.

\bibitem[Narendra and Thathachar(1974)]{narendra1974learning}
K.~S. Narendra and M.~A.~L. Thathachar.
\newblock Learning automata a survey.
\newblock \emph{IEEE Transactions on Systems, Man and Cybernetics}, 4:\penalty0
  323--334, 1974.

\bibitem[Newlund(2012)]{meetingNumber}
D.~Newlund.
\newblock Make your meetings worth everyone's time:
  \url{http://www.usatoday.com/USCP/PNI/Business/2012-06-20-PNI0620biz-career-%
getting-aheadPNIBrd\_ST\_U.htm}, 2012.

\bibitem[Olsen(2009)]{persuasivewords1}
H.~Olsen.
\newblock 108 persuasive words:
  \url{http://www.high-output.com/uncategorized/108-persuasive-words/}, 2009.

\bibitem[Panko and Kinney(1995)]{panko1995meeting}
R.~R. Panko and S.~T. Kinney.
\newblock Meeting profiles: Size, duration, and location.
\newblock In \emph{Proceedings of the Twenty-Eighth Hawaii International
  Conference on System Sciences}, volume~4, pages 1002--1011, 1995.

\bibitem[Purver et~al.(2007)Purver, Dowding, Niekrasz, Ehlen, Noorbaloochi, and
  Peters]{purver2007detecting}
M.~Purver, J.~Dowding, J.~Niekrasz, P.~Ehlen, S.~Noorbaloochi, and S.~Peters.
\newblock Detecting and summarizing action items in multi-party dialogue.
\newblock In \emph{Proceedings of the 8th Special Interest Group on Discourse
  and Dialogue Workshop on Discourse and Dialogue}, 2007.

\bibitem[Reiter and Rigoll(2005)]{reiter2005multimodal}
S.~Reiter and G.~Rigoll.
\newblock Multimodal meeting analysis by segmentation and classification of
  meeting events based on a higher level semantic approach.
\newblock In \emph{Proceedings of International Conference on Acoustics,
  Speech, and Signal Processing}, volume~2, pages 161--164, 2005.

\bibitem[Romano~Jr and Nunamaker~Jr(2001)]{romano2001meeting}
N.~C. Romano~Jr and J.~F. Nunamaker~Jr.
\newblock Meeting analysis: Findings from research and practice.
\newblock In \emph{Proceedings of the 34th Annual Hawaii International
  Conference on System Sciences}, 2001.

\bibitem[Ron et~al.(1996)Ron, Singer, and Tishby]{ron1996power}
D.~Ron, Y.~Singer, and N.~Tishby.
\newblock The power of amnesia: Learning probabilistic automata with variable
  memory length.
\newblock \emph{Machine learning}, 25\penalty0 (2-3):\penalty0 117--149, 1996.

\bibitem[Scheidel(1967)]{scheidel1967persuasive}
T.~M. Scheidel.
\newblock \emph{Persuasive Speaking.}
\newblock Scott, Foresman and Company, 1967.

\bibitem[Shahaf et~al.(2012)Shahaf, Guestrin, and Horvitz]{shahaf2012trains}
D.~Shahaf, C.~Guestrin, and E.~Horvitz.
\newblock Trains of thought: Generating information maps.
\newblock In \emph{Proceedings of the 21st International Conference on World
  Wide Web}, pages 899--908, 2012.

\bibitem[Sim and Park(2003)]{sim2003consensus}
J.~S. Sim and K.~Park.
\newblock The consensus string problem for a metric is {NP}-complete.
\newblock \emph{Journal of Discrete Algorithms}, 1\penalty0 (1):\penalty0
  111--117, 2003.

\bibitem[{SRI International}(2003-2009)]{calo}
{SRI International}.
\newblock {CALO}: Cognitive assistant that learns and organizes.
  \url{http://caloproject.sri.com}, 2003-2009.

\bibitem[Sternthal et~al.(1978)Sternthal, Phillips, and
  Dholakia]{sternthal1978persuasive}
B.~Sternthal, L.W. Phillips, and R.~Dholakia.
\newblock The persuasive effect of scarce credibility: a situational analysis.
\newblock \emph{Public Opinion Quarterly}, 42\penalty0 (3):\penalty0 285--314,
  1978.

\bibitem[Stolcke et~al.(2000)Stolcke, Ries, Coccaro, Shriberg, Bates, Jurafsky,
  Taylor, Martin, Ess-Dykema, and Meteer]{stolcke2000dialogue}
A.~Stolcke, K.~Ries, N.~Coccaro, E.~Shriberg, R.~Bates, D.~Jurafsky, P.~Taylor,
  R.~Martin, C.V. Ess-Dykema, and M.~Meteer.
\newblock Dialogue act modeling for automatic tagging and recognition of
  conversational speech.
\newblock \emph{Computational Linguistics}, 26\penalty0 (3):\penalty0 339--373,
  2000.

\bibitem[Tur et~al.(2010)Tur, Stolcke, Voss, Peters, Hakkani-Tur, Dowding,
  Favre, Fernandez, Frampton, Frandsen, Frederickson, Graciarena, Kintzing,
  Leveque, Mason, Niekrasz, Purver, Riedhammer, Shriberg, Tien, Vergyri, and
  Yang]{tur2008calo}
G.~Tur, A.~Stolcke, L.~Voss, S.~Peters, D.~Hakkani-Tur, J.~Dowding, B.~Favre,
  R.~Fernandez, M.~Frampton, M.~Frandsen, C.~Frederickson, M.~Graciarena,
  D.~Kintzing, K.~Leveque, S.~Mason, J.~Niekrasz, M.~Purver, K.~Riedhammer,
  E.~Shriberg, Jing Tien, D.~Vergyri, and Fan Yang.
\newblock The {CALO} meeting assistant system.
\newblock \emph{IEEE Transactions on Audio, Speech, and Language Processing},
  18\penalty0 (6):\penalty0 1601--1611, 2010.

\bibitem[van~der Aalst(1998)]{van1998application}
W.~M.~P. van~der Aalst.
\newblock The application of petri nets to workflow management.
\newblock \emph{Journal of Circuits, Systems, and Computers}, 8\penalty0
  (01):\penalty0 21--66, 1998.

\bibitem[Wang et~al.(2012)Wang, Agichtein, and Benzi]{wang2012tm}
Y.~Wang, E.~Agichtein, and M.~Benzi.
\newblock Tm-lda: efficient online modeling of latent topic transitions in
  social media.
\newblock In \emph{Proceedings of the 18th International Conference on
  Knowledge Discovery and Data Mining}, pages 123--131, 2012.

\bibitem[Wikipedia(2013)]{wikiAutomata}
Wikipedia.
\newblock Probabilistic automation--- {W}ikipedia{,} the free encyclopedia,
  2013.
\newblock URL \url{http://en.wikipedia.org/wiki/Probabilistic_automaton}.
\newblock [Online; accessed 28-May-2013].

\end{thebibliography}
\bibliographystyle{plainnat}

\newpage
\section*{Appendix}

\subsection*{Proof of Theorem 1.}
We will use Hoeffding's inequality combined with the union bound to create a uniform generalization bound over all viable templates. The main step in doing this is to count the number of possible viable templates. Let us do this now.

Let $\Lambda$  be the set of possible dialogue acts, and denote $|\Lambda|$
 as the number of elements in the set. We will calculate the number of templates that 
are of size less than or equal to $L$, which is the size of our function class in statistical learning theory. 

For a template of exactly length $n$, there are $|\Lambda|^n$ possible assignments of dialogue acts for the 
templates. Also, for a template of length $n$, there are at most ${n(n-1)/2 \choose B}$ possible 
assignments of $B$ backward arrows, where $B \leq n$. To see this, consider the set of backwards arrows as 
represented by an $n\times n$ adjacency matrix, where only the part below the diagonal could be 1. There are $n^2$ 
total elements in the matrix, $n$ on the diagonal, so $n(n-1)$ off diagonal elements, and $n(n-1)/2$ elements 
in the lower triangle. If exactly $b$ of these can be 1, the total number of possibilities is at most ${n(n-1)/2 \choose b}$. There 
can be up to $B$ backward arrows, so the total number of possibilities is at most
\[\sum_{b=0}^{\min(B,n)} {n(n-1)/2 \choose b}.\]
 Note that this number is an upper bound, as usually we cannot have more than one backwards arrow leaving or entering a node. Finally 
we could have $n$ anywhere between 0 and $L$ so the final number of possible templates has upper bound:
\[
\sum_{n=0}^{L} |\Lambda|^n \sum_{b=0}^{\min(B,n)} {n(n-1)/2 \choose b}.
\]
Hoeffding's inequality applies to arbitrary bounded loss functions. This, combined with a union bound over all viable templates, yields the statement of the theorem.

 \end{document}